\documentclass[12pt]{article}

\usepackage{setspace}
\usepackage{amsmath,amsfonts}
\usepackage{theorem}

\usepackage{epsfig}

\usepackage{xcolor}

\usepackage{lscape}

\usepackage{multirow}

\usepackage{rotating}

\usepackage{tikz}

\usepackage{makecell}
\newcolumntype{x}[1]{>{\centering\arraybackslash}p{#1}}
\usepackage[sort&compress]{natbib}
\bibpunct{(}{)}{,}{a}{,}{,}

\usepackage{algorithm}
\usepackage{algorithmic}
\usepackage{appendix}

\theoremstyle{plain} { \theorembodyfont{\rmfamily}

}
\newtheorem{proposition}{Proposition}

\def\qed{\quad Q.E.D.}

\newenvironment{pfof}[1]{\vspace{1ex}\noindent{\em Proof of #1}\hspace{0.5em}}
    {\vspace{1ex}}

\newcommand{\add}{\color{black}}

\begin{document}
\title{From Time-inconsistency to Time-consistency for Optimal Stopping Problems }
\author{Sang Hu\thanks{School of Data Science, The Chinese University of Hong Kong, Shenzhen, Guangdong Province, China 518172.
Email: \texttt{husang@cuhk.edu.cn}.}
\and Zihan Zhou\thanks{School of Data Science, The Chinese University of Hong Kong, Shenzhen, Guangdong Province, China 518172.
Email: \texttt{zihanzhou@link.cuhk.edu.cn}}
}

\maketitle

\abstract{For optimal stopping problems with time-inconsistent preference, we measure the inherent level of time-inconsistency by taking the time required to turn the naive strategies into the sophisticated ones. In particular, when in a repeated experiment, the naive agent can observe her actual sequence of actions that is inconsistent with what she has planned at the initial time, and she would then choose her immediate action based on the observations of her later actual behavior. The procedure is repeated until her actual sequence of actions is consistent with her plan at any time. We show that for the preference value of cumulative prospect theory, in which the time-inconsistency is due to the probability distortion, the higher the degree of probability distortion, the more severe the level of time-inconsistency, and the more time required to turn the naive strategies into the sophisticated ones.}

\vspace{1ex}


\section{Introduction}

Optimal stopping problems arise in many economic and financial decision-making scenarios -- for example, when an entrepreneur completes a project, an investor sells stock, or a gambler quits gambling at the casino.
The stopping decision depends on the preference of the entrepreneur, the investor, or the gambler.
In a dynamic decision process, the agent plans a sequence of actions to be taken at each time point. Then, it is possible that later when the agent revisits the same optimal stopping problem at a different time than the initial time, she finds that her previously planned sequence of actions is no longer optimal according to her current preference. This phenomenon is called {\em time-inconsistency}.

Time-inconsistency can be observed in many dynamic decision problems. In the casino gambling problem, for example, a gambler may first plan her strategy as follows: She would continue gambling if she gains and stop gambling if she loses -- termed the ``loss-exit'' strategy. When she does gambles, her actual behavior may be opposite to what she has planned: She stops gambling if she gains and continues gambling if she loses -- termed the ``gain-exit'' strategy. This is because her preference characterized by cumulative prospect theory leads to time-inconsistency. See \cite{Barberis2012:Casino}, \citet{HeHuOblojZhou2016:Randomization}, \cite{HeEtal2019:StoppingStrategies} and \cite{HuEtal2022:ACasinoGamblingModel} for more detailed discussions.
Other possible scenarios that induce time-inconsistency include mean-variance preference, state-dependent reference, or non-exponential discount factors.

\cite{BjordTMurgoci:08u} concludes that there are three approaches for dealing with time-inconsistency. First, the agent ignores the time-inconsistency issue and derives the optimal strategy at the initial time to be her strategy. This is called the pre-committed type. Second, the agent constructs a strategy such that her current decision is the best based on expectations on her future decisions. This is called the sophisticated type. Third, the agent continuously derives her current best strategy and deviates from what she has planned before. This is called the naive type. The classification of three types of agents parallels the one used in the literature on hyperbolic discounting, which is one reason for the time-inconsistency; see \cite{MachinaM1989} for more discussions on three types of agents.

{\add A large literature studies the behavior of three types of agents in various decision problems, but the connections between them and even possible transformations into each other remain unclear.}
\citet{HuangEtal2017:StoppingBehaviors} first prove that in a one-dimensional diffusion process when the payoff functional satisfies regularity conditions, the sophisticated equilibrium of the stopping problem can be obtained as a fixed point of an operator, which represents strategic reasoning that takes the future selves’ behaviors into account. In other words, the strategic reasoning may turn a naive agent into a sophisticated one.

In this work, we consider the time-inconsistent optimal stopping problem with randomization in a discrete-time setting. {\add In particular, we propose a measure of time-inconsistency by measuring the gap between the naive strategy and the sophisticated strategy.} In a finite time horizon, {\add we use the binomial tree to describe the underlying state process, where each node represents a pair of time and state.} The sophisticated strategies can be backward derived from the terminal time to the initial time, whereas naive strategies are derived by optimizing the stopping problem at each node. By taking the naive agent's actual behavior into consideration, the agent's strategies eventually match with sophisticated strategies after several rounds of training on strategic reasoning.
In the example of cumulative prospect theory preferences (CPT) {\add -- one of the well-known non-expected utility theories}, the higher the degree of probability distortion, the more severe the level of time-inconsistency, and the more time required to turn the naive strategies into the sophisticated ones; {\add details are shown in Section \ref{se:CPT}.

According to the time-inconsistency measure, we design an algorithm to transform the naive strategy into the sophisticated one. The algorithm can be applied to any time-inconsistent stopping problem. In addition to the cumulative prospect theory preferences -- where the time-inconsistency is due to the probability distortion -- we also consider the present-biased preferences -- where the time-inconsistency is due to the non-exponential discount factor. We derive analytical results on how many rounds are needed to achieve the sophisticated strategy from the naive one in the optimal stopping problem with immediate cost and with immediate reward. The transformation on the stopping strategies can also be made in other types of time-inconsistent stopping problem.

In addition to the above mentioned literature, our work is also related to the following literature on time-inconsistent optimal stopping problem with cumulative prospect theory:  \cite{XuZhou2012:OptimalStoppingunderProbabilityDistortion},  \cite{EbertStrack2012:UntilTheBitterEnd},  \citet{EbertStrack2016:NeverEverGettingStarted}, \citet{HendersonHobsonTse14}, \citet{BelomestnyKratschmer2017:OptimalStopping}, \citet{HendersonHobosonTse2018probability}. The key difference is that these literatures study the optimal stopping problem in continuous time, while our focus is on discrete time. There is also extensive literature on general optimal stopping problem, e.g., \cite{DayanikKaratza2003}, \cite{Shiryaev2007}, etc. The following literature discusses the general time-inconsistent decision problem: \citet{Strotz1955:MyopiaInconsistency}, \citet{EkelandLazrak2006:BeingSeriousAboutNonCommitment}, \citet{EkelandPirvu2008:InvestConsptnWithoutCommit}, \citet{EbertEtal2017:Discounting}, \citet{TanEtal2021:FailureOfSmoothPasting}, \citet{Christensen2018finding}, \citet{ChristensenLindensjo2020:OnTimeInconsistentStopping}, \citet{HuangNguyenHuu2018:TimeConsistent}, \citet{HuangYu2021:OptimalStopping}, \citet{HeZhou2022:WhoAreI}, etc.

}

The rest of this paper is organized as follows. In section \ref{se:Model}, we introduce the model of the optimal stopping problem with randomization, time-inconsistent preferences, and the different types of agents. In section \ref{se:Measure}, we establish iterations to turn the naive strategies into the sophisticated ones as a measure of time-inconsistency. We illustrate the iteration procedure in section \ref{se:CPT}, using CPT as an example of time-inconsistent preferences. {\add Section \ref{se:Presentbiased} presents the analytical results of transforming the time-inconsistent strategies into time-consistent ones under the present-biased preferences.} Section \ref{se:Conclusion} provides the conclusion.

\section{Model}\label{se:Model}

\subsection{Optimal stopping}
Consider an optimal stopping problem faced by an agent in a discrete-time simple symmetric random walk. Let $\Delta_{0,0}$ stand for the set of all feasible time-state pairs in the simple symmetric random walk up to time $T$. Consider Markovian stopping strategies with external randomization allowed. At time 0 with initial state 0, the agent determines her stopping strategy by choosing a sequence of actions ${\bf a}_{0,0} = \{a_{0,0}(t,x)\}_{(t,x) \in \Delta_{0,0}}$, where $a_{0,0}(t,x) \in [0,1]$ stands for the probability to stop at node $(t,x)$. If $a_{0,0}(t,x) = p$ for some $p \in [0,1]$, the agent tosses a (biased) coin to determine whether to stop or not. If the coin lands on heads, the agent chooses to continue; otherwise, she chooses to stop. The probability of tossing a tail is equal to $p$. In particular, if $a_{0,0}(t,x) = 1$, the agent chooses to stop for sure at node $(t,x)$; if $a_{0,0}(t,x) = 0$, she chooses to continue for sure at node $(t,x)$. {\add See Figure \ref{fig_tree} for illustration.

\begin{figure}[!htbp]
  \centering
  \begin{minipage}[t]{0.33\textwidth}
  \centering
  \includegraphics[width=\textwidth]{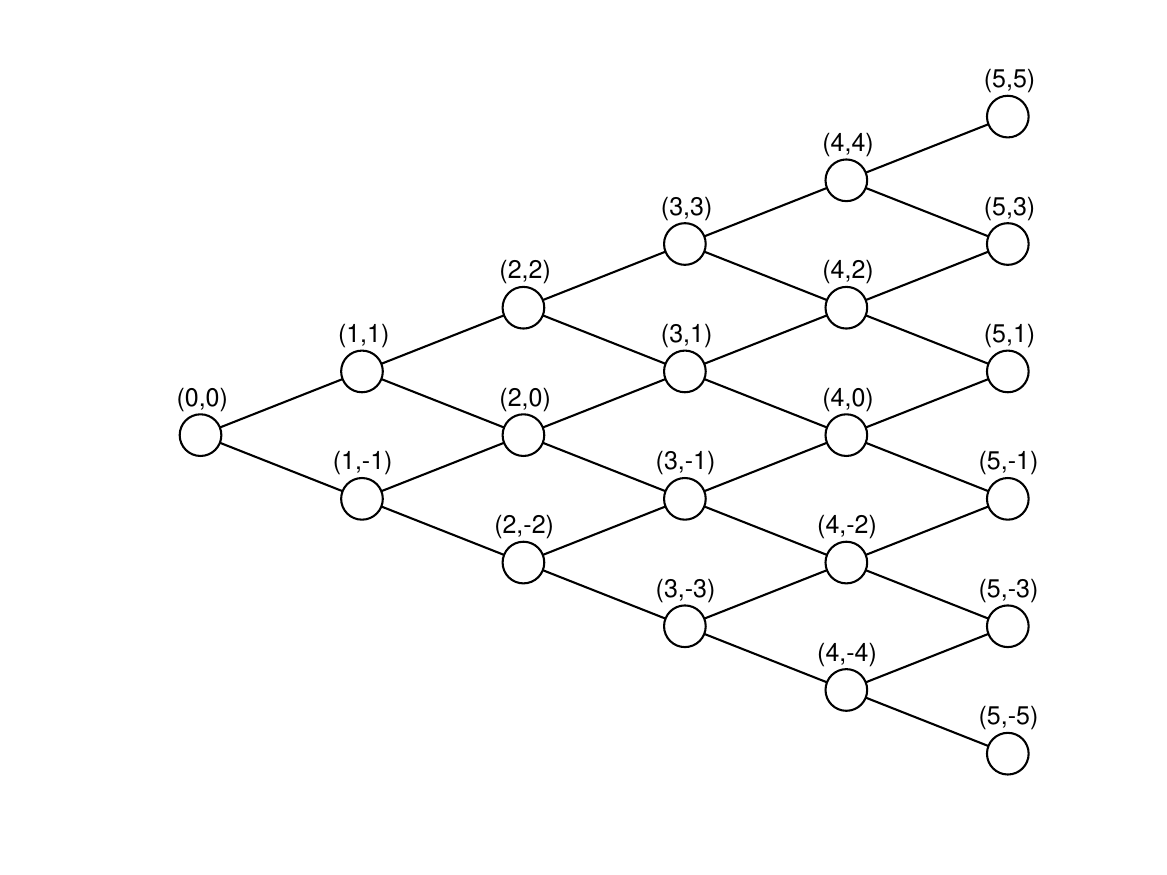}\\
  \end{minipage}
  \caption{A five-period binomial tree. The pair above each node stands for the current time and state.}\label{fig_tree}
\end{figure}

}

Suppose that the agent tosses a {\add head} at time 0, which means that she would continue. Then, at time 1 and state $j$, where $j \in \{-1,1\}$, let $\Delta_{1,j}$ stand for the set of all the remaining feasible time-state pairs in simple symmetric random walk after time 1 and starting at state $j$, up to time $T$. Suppose that the agent revisits the problem to choose her action, regardless of $a_{0,0}(1,j)$ -- the plan she made at time 0 for node $(1,j)$. Again, at time 1 she chooses a sequence of actions denoted by ${\bf a}_{1,j} = \{a_{1,j}(t,x)\}_{(t,x) \in \Delta, t \ge 1}$, where $a_{1,j}(t,x) \in [0,1]$ stands for the probability to stop at node $(t,x)$ according to the plan made at time 1. This pattern continues until the terminal time $T > 0$ if the agent has not stopped yet.

\subsection{Time-inconsistent preferences}
Let $V_{t,x}({\bf a})$ denote the preference value of the agent at time $t$ and state $x$ when applying a sequence of actions ${\bf a} = \{a(t,x)\}$ afterward.
The preference value function $V$ is called {\em time-inconsistent} if there exist $(t,x)$ and $(t',x')$ with $t < t'$ and $(s,y) \in \Delta_{t,x} \cap \Delta_{t',x'}$ such that
\begin{align*}
a^*_{t,x}(s,y) \neq a^*_{t',x'}(s,y),
\end{align*}
where ${\bf a}^*_{t,x} = \{a^*_{t,x}(s,y)\}$ is the optimal action sequence planned at $(t,x)$, that is, 
\begin{align*}
\begin{array}{cl}
{\bf a}^*_{t,x} := \arg\max_{{\bf a} \in \mathbb{R}^{|\Delta_{t,x}|}([0,1])} V_{t,x}({\bf a}),
\end{array}
\end{align*}
where $\mathbb{R}^n([0,1])$ stands for the $n$-dimensional vector taking values in $[0,1]$ in a point-wise manner, and $|A|$ is the number of elements in set $A$.
In other words, as long as there exist certain differences in the actions between those that are planned at different times, the preference value $V$ is time-inconsistent. It is straightforward to verify that the total number of nodes in the set $\Delta_{t,x}$ is $(T-t+1)(T-t+2)/2$.
Note that there exists a one-to-one correspondence between the elements in the $|\Delta_{t,x}|$-dimensional vector ${\bf a}$ to the action $a(s,y)$ taken at node $(s,y)$ for some value $s \ge t$ and $y \in \{-s,-(s-2),...s-2,s\}$. In particular, with current time $t$ and state $x$, the $j$-th element in the vector ${\bf a}$ corresponds to $a_{t,x}(s,y)$, where $s$ is such that $(s-t)(s-t+1)/2 < j \le (s-t+1)(s-t+2)/2$ and $y = s - 2j+(s-t)(s-t+1)+2$.
Hence, we do not differentiate the vector ${\bf a}$ and the action sequence $\{a(t,x)\}$ in the following.

Many situations arise in which the preference value is time-inconsistent. For example, if the preference is mean-variance, it is time-inconsistent because the variance term is nonlinear in the probability distributions. It may also be due to the non-exponential discount factor, which makes the preference time-inconsistent. On the other hand, some behavioral preference such as cumulative prospect theory has a probability distortion component in its preference value. This makes the preference value nonlinear in the probability distributions; therefore, the CPT preference is time-inconsistent.

\subsection{Agents}
Regarding the time-inconsistency issue, naturally, one may raise the question of which plan is going to be adopted by the agent. According to whether the agent is aware of time-inconsistency and whether the agent can commit herself to a predetermined plan, we classify the agents into three categories: naive, sophisticated, and pre-committed.

\subsubsection{Naive agent}
A {\em naive} agent does not realize the time-inconsistency. At any given time the naive agent just seeks an optimal solution at that moment, but she is only able to implement this solution at that moment. In other words, the naive agent continuously deviates from the plans that were made by herself previously. Her actual stopping strategies are called naive strategies.

At time 0, the naive agent determines her plan at that time to be ${\bf a}^*_{0,0}$ that solves
\begin{align*}
\begin{array}{cl}
\max_{{\bf a} \in \mathbb{R}^{|\Delta_{0,0}|}([0,1])} & V_{0,0}({\bf a}) .
\end{array}
\end{align*}
She then takes action $a^*_{0,0}(0,0)$. If she does not stop at time 0 and continues to time 1, then at time 1 with state $j \in \{-1,1\}$, she disregards $a^*_{0,0}(1,j)$ -- the action she should take at that time according to the optimal plan ${\bf a}^*_{0,0}$ made at time $0$. Instead, she determines her strategy at that time to be ${\bf a}^*_{1,j}$ that solves
\begin{align*}
\begin{array}{cl}
\max_{{\bf a} \in \mathbb{R}^{|\Delta_{1,j}|}([0,1])} & V_{1,j}({\bf a}) .
\end{array}
\end{align*}
Then, her actual action taken at time $1$ with state $j$ is $a^*_{1,j}(1,j)$. Under a time-inconsistent preference, $a^*_{1,j}(1,j)$ may be completely different from $a^*_{0,0}(1,j)$.
The naive agent continuously seeks the optimal solution to be her action taken at time $t$ until terminal time $T$ if she has not stopped yet. Her actual strategy is as follows:
\begin{align*}
{\bf a}^{N(0)} = \Big(a^*_{0,0}(0,0),... a^*_{t,X_t}(t,X_t),...a^*_{T,X_T}(T,X_T)\Big),
\end{align*}
where $X_t \in \{-t, -(t-2),...t-2,t\}$ is the state variable at time $t$.

\subsubsection{Sophisticated agent}
A {\em sophisticated} agent realizes the time-inconsistency but has no commitment to any predetermined plan.
Hence, the sophisticated agent chooses consistent planning in the sense that she optimizes today by expecting her actions in the future.
The agent's selves at different times are considered to be the players of a game, and a consistent plan becomes {\em an intra-personal equilibrium} of the game, from which no selves are willing to deviate.
Let $\tilde {\bf a}_{T,X_T}$ be the strategy determined by the agent at time $T$ with state $X_T$. Note that the agent can only choose to stop at time $T$, that is,
\begin{align*}
\begin{array}{cl}
\tilde a_{T,X_T}(T,X_T) = 1.
\end{array}
\end{align*}
Then, at time $T-1$, the sophisticated agent determines her action taken at time $T-1$ according to the action taken at time $T$. Denote her plan at time $T-1$ to be $\tilde {\bf a}_{T-1,X_{T-1}}$, which solves
\begin{align*}
\begin{array}{cl}
\max_{{\bf a} \in \mathbb{R}^{|\Delta_{T-1,X_{T-1}}|}([0,1])} & V_{T-1, X_{T-1}}({\bf a}), \\
\text{subject to } & a(T,X_T) = \tilde a_{T,X_T}(T,X_T).
\end{array}
\end{align*}
Compared with the decision made by a naive agent at time $T-1$, it is the same because both essentially choose the best action at time $T-1$, which is a single-period problem.
The situation is different at time $T-2$. The sophisticated agent determines her action taken at time $T-2$ according to the action taken at time $T-1$ and $T$. Denote her plan at time $T-2$ to be $\tilde {\bf a}_{T-2,X_{T-2}}$, which solves
\begin{align*}
\begin{array}{cl}
\max_{{\bf a} \in \mathbb{R}^{|\Delta_{T-2,X_{T-2}}|}([0,1])} & V_{T-2, X_{T-2}}({\bf a}), \\
\text{subject to } & a(T,X_T) = \tilde a_{T,X_T}(T,X_T), \\
& a(T-1,X_{T-1}) = \tilde a_{T-1,X_{T-1}}(T-1,X_{T-1}).
\end{array}
\end{align*}
Compared with the decision made by a naive agent at time $T-2$, there exists an additional constraint in that
\begin{align*}
a(T-1,X_{T-1}) = \tilde a_{T-1,X_{T-1}}(T-1,X_{T-1}).
\end{align*}
The sophisticated agent determines her strategy sequentially until time 0 in the same fashion. Denote her plan at time $t$ to be $\tilde {\bf a}_{t,X_t}$, which solves
\begin{align*}
\begin{array}{cl}
\max_{{\bf a} \in \mathbb{R}^{|\Delta_{t,X_t}|}([0,1])} & V_{t,X_t}({\bf a}), \\
\text{subject to } & a(s, X_{s}) = \tilde a_{s,X_{s}}(s,X_{s}), \; s = t+1,...T.
\end{array}
\end{align*}
Consequently, the sophisticated agent's plan at any time is consistent with her actual strategy which is
\begin{align*}
{\bf a}^S = \Big(\tilde a_{0,0}(0,0),...\tilde a_{t,X_t}(t,X_t),...\tilde a_{T,X_T}(T,X_T)\Big).
\end{align*}

\subsubsection{Pre-committed agent}
The {\em pre-committed} agent can commit to her predetermined plan that is made at time 0, although her preference is time-inconsistent. Hence, her actual stopping strategy is consistent with her plan, which is called a pre-committed strategy. It is simply the optimal solution that maximizes the preference value $V_{0,0}$ at time 0 with the initial state 0, which is same as the problem faced by the naive agent at time 0:
\begin{align*}
\begin{array}{cl}
{\bf a}^*_{0,0} = \arg\max_{{\bf a} \in \mathbb{R}^{|\Delta_{0,0}|}([0,1])} V_{0,0}({\bf a}) .
\end{array}
\end{align*}
The pre-committed strategy is
\begin{align*}
{\bf a}^P = {\bf a}^*_{0,0} = \Big(a^*_{0,0}(0,0),...a^*_{0,0}(t,X_t),...a^*_{0,0}(T,X_T)\Big).
\end{align*}
Since the pre-committed strategy can be solved when solving naive strategy, in the following, we focus on the naive and sophisticated strategies.

\section{Turn to Time-consistency}\label{se:Measure}

In this section, we provide an algorithm to ``train'' a naive agent into a sophisticated one as a measure of time-inconsistency. Unlike a pre-committed agent, both the naive and sophisticated agents have no commitment device. To achieve a consistent plan, a naive agent needs to be trained to realize the time-inconsistency and modify her decisions accordingly.

Suppose that in a repeated experiment, the naive agent can observe her actual stopping behavior, based on which she realizes that her optimal plan made at any time is deviated by her decisions made at future time points. Then at time 0 when she plans a sequence of actions taken in a simple symmetric random walk up to time $T$, she (possibly incorrectly) anticipates that her future selves are going to adopt a certain strategy which is consistent with her observation on her actual stopping strategy. This anticipation changes her optimal solution at time 0 to be ${\bf a}^{(1)}_{0,0}$, which solves
\begin{align*}
\begin{array}{cl}
\max_{{\bf a} \in \mathbb{R}^{|\Delta_{0,0}|}([0,1])} & V_{0,0}({\bf a}), \\
\text{subject to } &  a(s,X_s) = a^{N(0)}(s,X_s), \; s = 1,...T.
\end{array}
\end{align*}
Note that compared with the agent without training at time $0$, there exist additional constraints such that
\begin{align*}
a(s,X_s) = a^{N(0)}(s,X_s), \; s = 1,...T.
\end{align*}

Suppose that the agent does not stop at time 0 and continues to time 1. Then she revisits the problem by the same logic: she anticipates that her future selves are going to adopt the actual stopping strategy and then chooses her optimal solution at time 1 with state $j \in \{-1,1\}$ accordingly. Denote her plan at time $1$ to be ${\bf a}^{(1)}_{1,j}$, which solves
\begin{align*}
\begin{array}{cl}
\max_{{\bf a} \in \mathbb{R}^{|\Delta_{1,j}|}([0,1])} & V_{1,j}({\bf a}), \\
\text{subject to } & a(s,X_s) = a^{N(0)}(s,X_s), \; s = 2,...T .
\end{array}
\end{align*}
At time $t$ with state $X_t$, she plans her action to be taken according to her actual stopping strategy afterwards. Denote her plan at time $t$ to be ${\bf a}^{(1)}_{t,X_t}$, which solves
\begin{align*}
\begin{array}{cl}
\max_{{\bf a} \in \mathbb{R}^{|\Delta_{t,X_t}|}([0,1])} & V_{t,X_t}({\bf a}), \\
\text{subject to } & a(s,X_{s}) = a^{N(0)}(s,X_{s}), \; s = t + 1,...T.
\end{array}
\end{align*}
After one round of training, the naive strategy becomes
\begin{align*}
{\bf a}^{N(1)} = \Big(a^{(1)}_{0,0}(0,0),... a^{(1)}_{t,X_t}(t,X_t),...a^{(1)}_{T,X_T}(T,X_T)\Big).
\end{align*}

If ${\bf a}^{N(1)} = {\bf a}^S$, then the naive agent has been successfully trained into a sophisticated one with consistent strategies. If ${\bf a}^{N(1)} \neq {\bf a}^S$, then the naive agent is still not fully sophisticated and requires more rounds of training. In other words, suppose that after $k$ rounds of training, the naive agent's actual strategy does not equal the sophisticated agent's strategy, that is, ${\bf a}^{N(k)} \neq {\bf a}^S$, $k \ge 1$. Then, at time $t$ with state $X_t$, the naive agent at the $(k+1)$-th round plans her action to be taken according to her actual $k$-th round of stopping strategy. Denote her $(k+1)$-th round's plan at time $t$ to be ${\bf a}^{(k+1)}_{t,X_t}$, which solves
\begin{align*}
\begin{array}{cl}
\max_{{\bf a} \in \mathbb{R}^{|\Delta_{t,X_t}|}([0,1])} & V_{t,X_t}(a), \\
\text{subject to } & a(s,X_{s}) = a^{N(k)}(s,X_{s}), \; s = t+1,...T.
\end{array}
\end{align*}
The constraints $a(s,X_{s}) = a^{N(k)}(s,X_{s})$ show that the agent anticipates that at time $s \ge t+1$, she is going to behave according to $k$-th stopping strategy.

In a $T$-period time horizon as above, the naive strategies are turned into the sophisticated ones after $T-1$ rounds of training at most, because the naive strategies are closer to the sophisticated ones by at least one time step after one round of training. The following proposition presents this result.
\begin{proposition}\label{prop:Tround}
Consider a $T$-period binomial tree. The naive strategies are the same as the sophisticated ones after $T-1$ rounds of training, that is, ${\bf a}^{N(T-1)} = {\bf a}^{S}$.
\end{proposition}

Indeed, one may need less than $T-1$ rounds to turn the naive strategies into the sophisticated ones. Once the naive agent's actual stopping strategy is the same as the sophisticated strategy, no more training is needed because the naive agent's actual stopping strategy is consistent with her plan eventually. The total number of rounds hence measures the level of time-inconsistency. The more the number of rounds needed, the higher the level of time-inconsistency. The iteration steps are summarized in the following algorithm.

\begin{center}
\begin{algorithm}[!htb]
\caption{From the naive strategies to the sophisticated strategies in a $T$-horizon binomial tree}
\begin{algorithmic}[1]
\FOR{each $t \in [0,T] \cap \mathbb{Z}$}
\FOR{each $j \in [1,t+1] \cap \mathbb{Z}$}
\STATE $x = t-2*(j-1)$;
\STATE optimize $V_{t,x}({\bf a})$;
\STATE take the optimal solution at node $(t,x)$ to be the action $a^{N(0)}(t,x)$;
\ENDFOR
\ENDFOR
\FOR{each $l \in [0,T] \cap \mathbb{Z}$}
\STATE $t = T-l$;
\FOR{each $j \in [1,t+1] \cap \mathbb{Z}$}
\STATE $x = t-2*(j-1)$;
\STATE optimize $V_{t,x}({\bf a})$, subject to $a(s,y) = a^{S}(s,y)$, $s \in [t+1,T] \cap \mathbb{Z}$, $y = s - 2*(b-1)$ for $b \in [1,s+1] \cap \mathbb{Z}$;
\STATE take the optimal solution at node $(t,x)$ to be the action $a^{S}(t,x)$;
\ENDFOR
\ENDFOR
\STATE $k = 1$;
\WHILE{${\bf a}^{N(k-1)} \neq {\bf a}^{S}$}
\FOR{each $t \in [0,T] \cap \mathbb{Z}$}
\FOR{each $j \in [1,t+1] \cap \mathbb{Z}$}
\STATE $x = t-2*(j-1)$;
\STATE optimize $V_{t,x}({\bf a})$, subject to $a(s,y) = a^{N(k-1)}(s,y)$, $s \in [t+1,T] \cap \mathbb{Z}$, $y = s - 2*(b-1)$ for $b \in [1,s+1] \cap \mathbb{Z}$;
\STATE take the optimal solution at node $(t,x)$ to be the action $a^{N(k)}(t,x)$;
\ENDFOR
\ENDFOR
\STATE $k = k+1$;
\ENDWHILE
\end{algorithmic}
\end{algorithm}
\end{center}

\newpage

\section{CPT Preferences}\label{se:CPT}

In this section we apply the measure of time-inconsistency defined in previous section to a specific optimal stopping problem. In particular, we consider the cumulative prospect theory of \cite{TverskyKahneman1992:CPT}, which is a time-inconsistent preference because of the probability distortion in the preference value.

\subsection{Cumulative prospect theory}

The expected utility (EU) framework in classical economics theories has been challenged by more and more empirical evidence. As an alternative to EU, the cumulative prospect theory (CPT) proposed by \citet{TverskyKahneman1992:CPT} is one of the well-known non-expected utility theories and has been widely studied in recent years.
{\add Cumulative prospect theory can accommodate both risk-averse and risk-seeking behaviors, which are difficult to reconcile in classical expected utility framework, thus providing new explanations for many well-known empirical puzzles, such as the disposition effect \citep{ShefrinStatman1985:Disposition,OdeanT:98de} and the equity premium puzzle \citep{MehraRPrescottE:85ep,BenartziSThalerR:95mla,BarberisNHuangM:01ma}.}
In this subsection, we briefly review the cumulative prospect theory.

In evaluating uncertain payoffs according to CPT, there are four important features that differentiate CPT from the traditional EU. First, there exists a reference point in the utility function $u(\cdot)$. The values above the reference point are called gains and those below the reference point are losses. In CPT the values applied to the utility function are gains and losses, rather than the total wealth level in EU. Second, there is one diminishing sensitivity in both gains and losses. It then implies an S-shaped utility function: in gains, the utility function exhibits concavity and in losses, convexity.
Third, for the same magnitude of gains and losses, one is more sensitive to the disutility of losses compared with the utility of gains, which is termed loss aversion. \citet{TverskyKahneman1992:CPT} proposed an analytical form of such an S-shaped utility $u(\cdot)$:
\begin{align}\label{eq:utility}
u(x) =
\begin{cases}
(x-B)^{\alpha_+} &, \text{ if } x \geq B\\
-\lambda (B-x)^{\alpha_-} &, \text{ if } x < B,
\end{cases}
\end{align}
where $0 < \alpha_\pm < 1$ signifies that $u(\cdot)$ is concave in the gain domain and convex in the loss one, and $\lambda > 1$ is the degree of loss aversion.

Forth, the probability weighting functions $w_\pm(\cdot)$ are applied in the preference evaluation. The existence of probability weighting makes the evaluation of risky payoff nonlinear in probability distribution because one does not use objective probabilities to evaluate events. An inverse-S shaped probability weighting function is concave in the lower-left corner for small probabilities close to 0 and convex in the upper-right corner for large probabilities close to 1. Note that $w_+$ is applied to gains, and $w_-$ is applied to losses. Then inverse S-shaped probability weighting functions lead to the effect that events with small probabilities of either large gains or losses are overweighted, events with large probabilities of small gains or losses are overweighted, and events with moderate probabilities of moderate gains or losses are underweighted. \citet{TverskyKahneman1992:CPT} suggested an analytical form of $w_\pm(\cdot)$, which is inverse S-shaped:
\begin{align}\label{eq:distortion}
w_\pm(p) = \frac{p^{\delta_\pm}}{(p^{\delta_\pm}+(1-p)^{\delta_\pm})^{\frac{1}{\delta_\pm}}}, \; p \in [0,1],
\end{align}
where $\delta_\pm \in (0.278,1)$ are the degrees of probability distortion in gains and losses, while $\delta_\pm = 1$ simply degenerates to no distortion.

Suppose that a sequence of actions $\{a(t,x)\}_{(t,x) \in \Delta_{0,0}}$ is applied in the simple symmetric random walk. Let $p_{0,0}^{{\bf a}}(n)$ be the probability of achieving state $n$, starting from $(0,0)$, with a strategy ${\bf a}$, $n = -T,...,-1,0,1,...T$.
Suppose that the reference point is the initial state 0. Then at time 0 the CPT preference value for this sequence of action $\{a(t,x)\}_{(t,x) \in \Delta_{0,0}}$ is
\begin{equation}\label{eq:CPTvalue}
\begin{array}{cl}
V_{0,0}({\bf a}) = & \sum_{n=1}^{T} u(n) \Big(w_+\left(\sum_{j=n}^{\infty} p_{0,0}^{{\bf a}}(j)\right)-w_+\left(\sum_{j=n+1}^{\infty} p_{0,0}^{{\bf a}}(j)\right)\Big) \\
& +  \sum_{n=1}^{T} u(-n) \Big(w_- \left(\sum_{j=n}^{\infty} p_{0,0}^{{\bf a}}(-j)\right)-w_-\left(\sum_{j=n+1}^{\infty} p_{0,0}^{{\bf a}}(-j)\right)\Big).
\end{array}
\end{equation}

Note that the CPT preference is time-inconsistent because of the probability distortion. From the mathematical point of view, $V$ is nonlinear in neither $p(n)$ nor the cumulative probabilities $\sum_{n^T} p(n)$. The agent at different time points evaluates the same event with inconsistent probability weights since the probabilities are distorted differently.
In particular, at time $t = 1,...T-1$ with state $x \in \{-t, -(t-2),...,(t-2),t\}$, the preference value of applying a sequence of actions $\{a(s,y)\}_{(s,y) \in \Delta_{t,x}}$ is
\begin{equation}\label{eq:CPTvalue}
\begin{array}{cl}
V_{t,x}({\bf a}) = & \sum_{n=1}^{T} u(n) \Big(w_+\left(\sum_{j=n}^{\infty} p_{t,x}^{{\bf a}}(j)\right)-w_+\left(\sum_{j=n+1}^{\infty} p_{t,x}^{{\bf a}}(j)\right)\Big) \\
& +  \sum_{n=1}^{T} u(-n) \Big(w_- \left(\sum_{j=n}^{\infty} p_{t,x}^{{\bf a}}(-j)\right)-w_-\left(\sum_{j=n+1}^{\infty} p_{t,x}^{{\bf a}}(-j)\right)\Big),
\end{array}
\end{equation}
where $p_{t,x}^{{\bf a}}(n)$ stands for the probability of achieving state $n$, starting from node $(t,x)$, with strategy ${\bf a}$, $n = -T,...-1, 0, 1,...T$.

\subsection{Five-period example}

We show in this subsection the procedure of training the naive strategies into the sophisticated ones through numerical examples in a five-period binomial tree.
Recall the utility function \eqref{eq:utility} and probability weighting function \eqref{eq:distortion}.

\begin{figure}[!htbp]
  \centering
  \begin{minipage}[t]{0.327\textwidth}
  \centering
  \includegraphics[width=\textwidth]{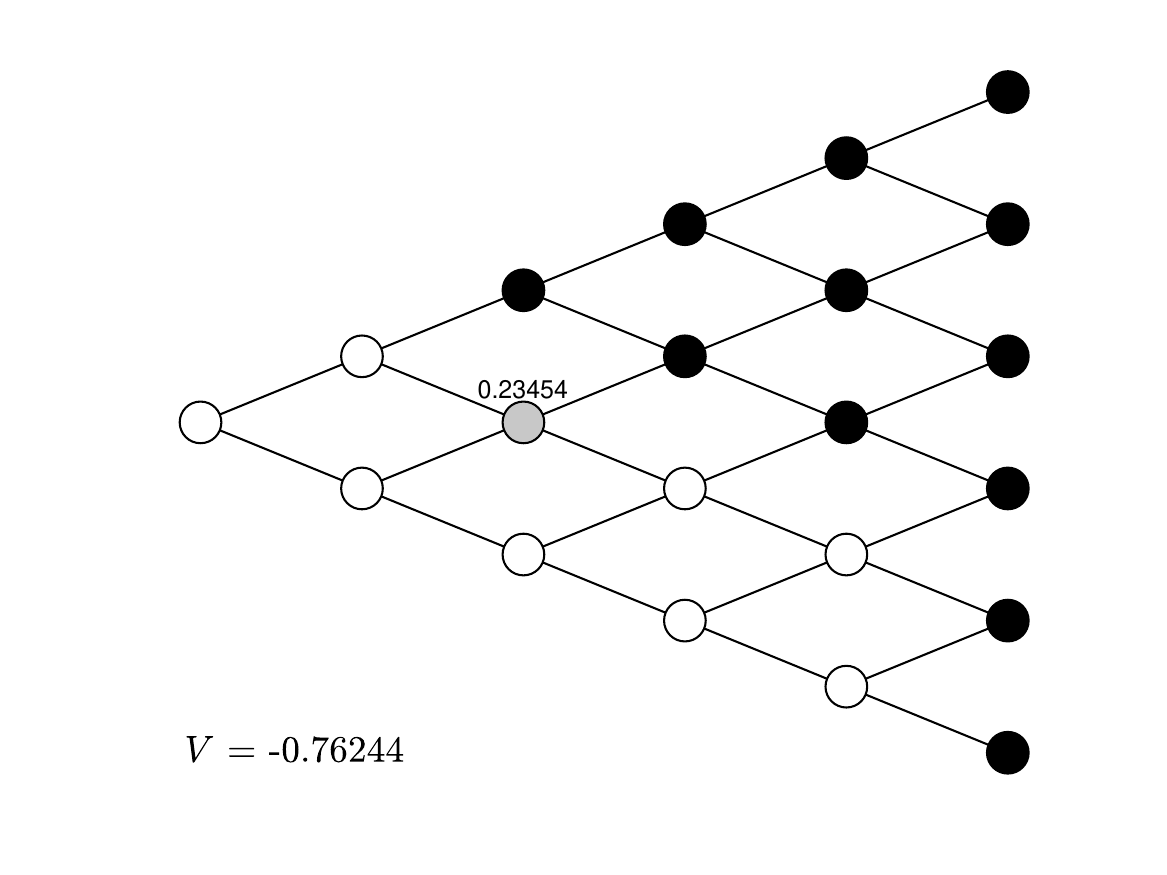}\\
  \end{minipage}
  \begin{minipage}[t]{0.327\textwidth}
  \centering
  \includegraphics[width=\textwidth]{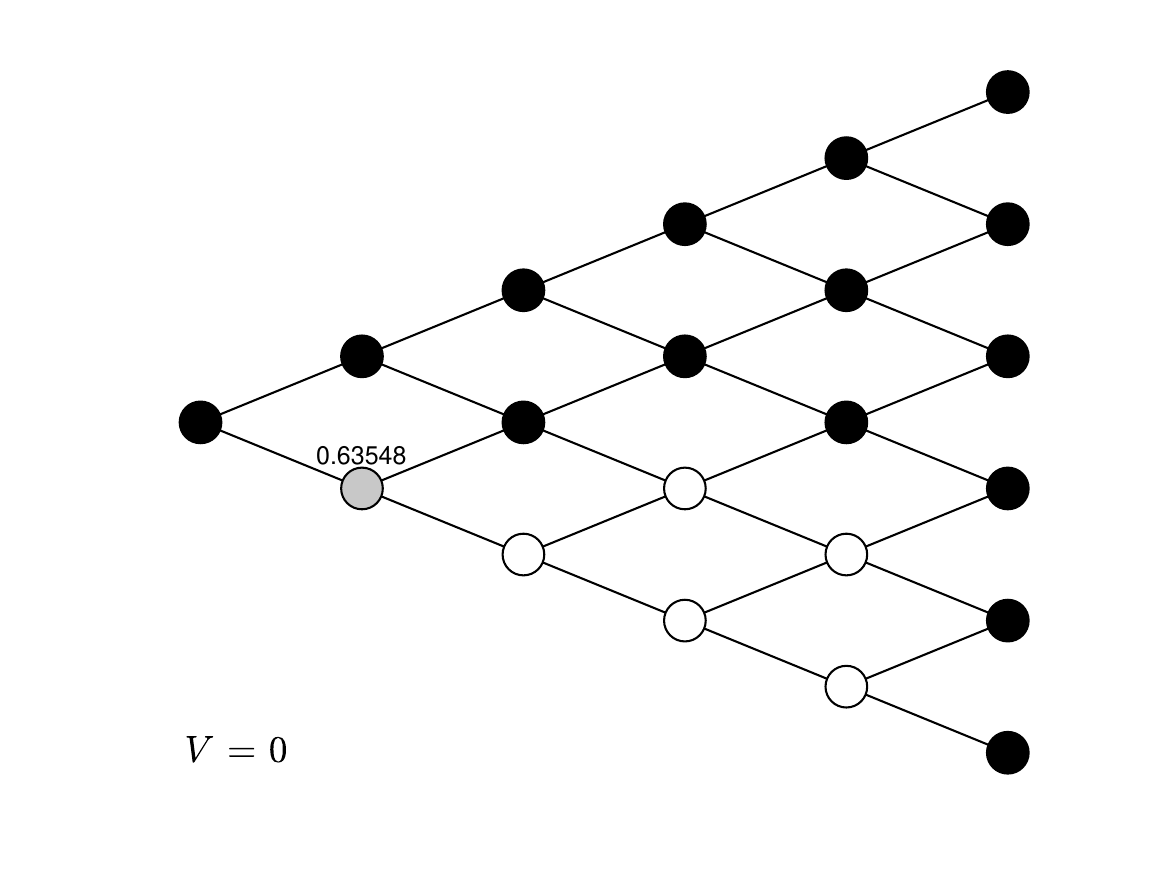}\\
  \end{minipage}
  \begin{minipage}[t]{0.327\textwidth}
  \centering
  \includegraphics[width=\textwidth]{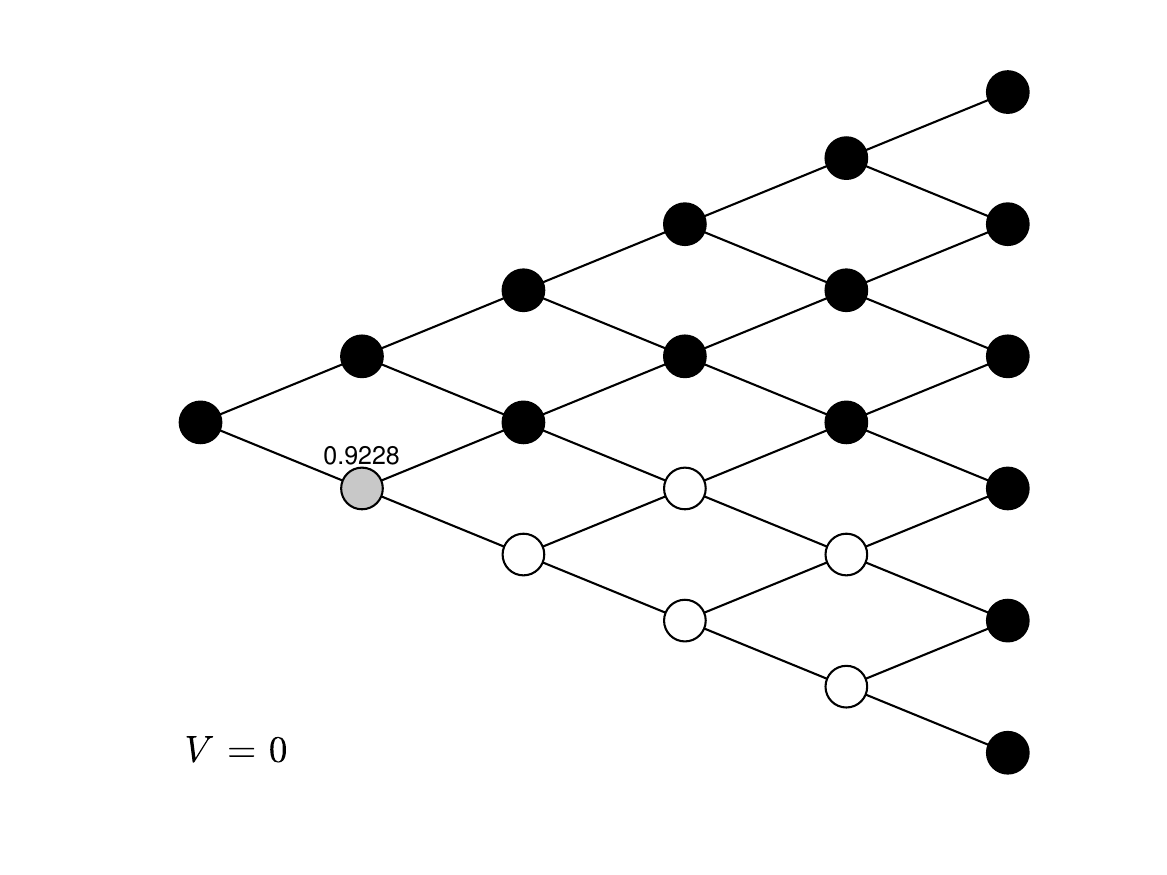}\\
  \end{minipage}
  \caption{Let $\alpha_\pm = 0.9$, $\delta_\pm = 0.5$, $\lambda = 1.5$. After two rounds the naive strategy is turned into the sophisticated strategy. The black nodes stand for stopping, the white nodes stand for continuing, and the grey nodes stand for randomization with the number above being the probability to stop.}\label{fig1}
\end{figure}

First, let $\alpha_\pm = 0.9$, $\delta_\pm = 0.5$, $\lambda = 1.5$. The first graph of Figure \ref{fig1} shows the naive strategy in a five-period binomial tree, which is essentially stopping in gains except node $(1,1)$, continuing in losses, and taking randomization at node $(2,0)$ with probability to stop equal to 0.23454. The initial action is continuing at node $(0,0)$. After observing such actual behavior and taking the subsequent actions into consideration, the naive agent updates her strategy at each node, as shown in the second graph in Figure \ref{fig1}. The action at node $(0,0)$ is stopping, in sharp contrast to the previous one. After one more round of training the naive strategy becomes exactly the same as the sophisticated one shown in the third graph of Figure \ref{fig1}. In other words, the naive strategy is trained into the sophisticated one after two rounds. The corresponding objective value as characterized by CPT increases as the naive strategy approaches the sophisticated one.

\begin{figure}[!htbp]
  \centering
  \begin{minipage}[t]{0.33\textwidth}
  \centering
  \includegraphics[width=\textwidth]{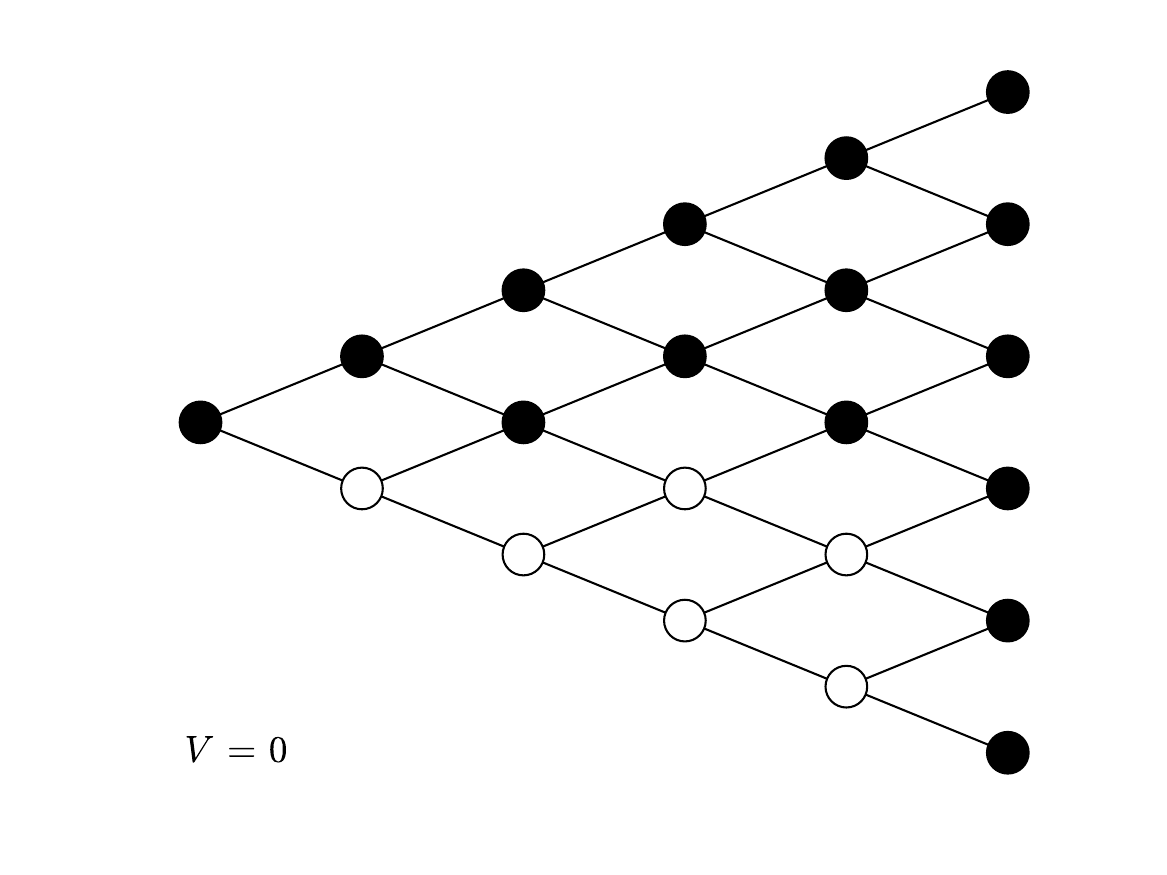}\\
  \end{minipage}
  \begin{minipage}[t]{0.33\textwidth}
  \centering
  \includegraphics[width=\textwidth]{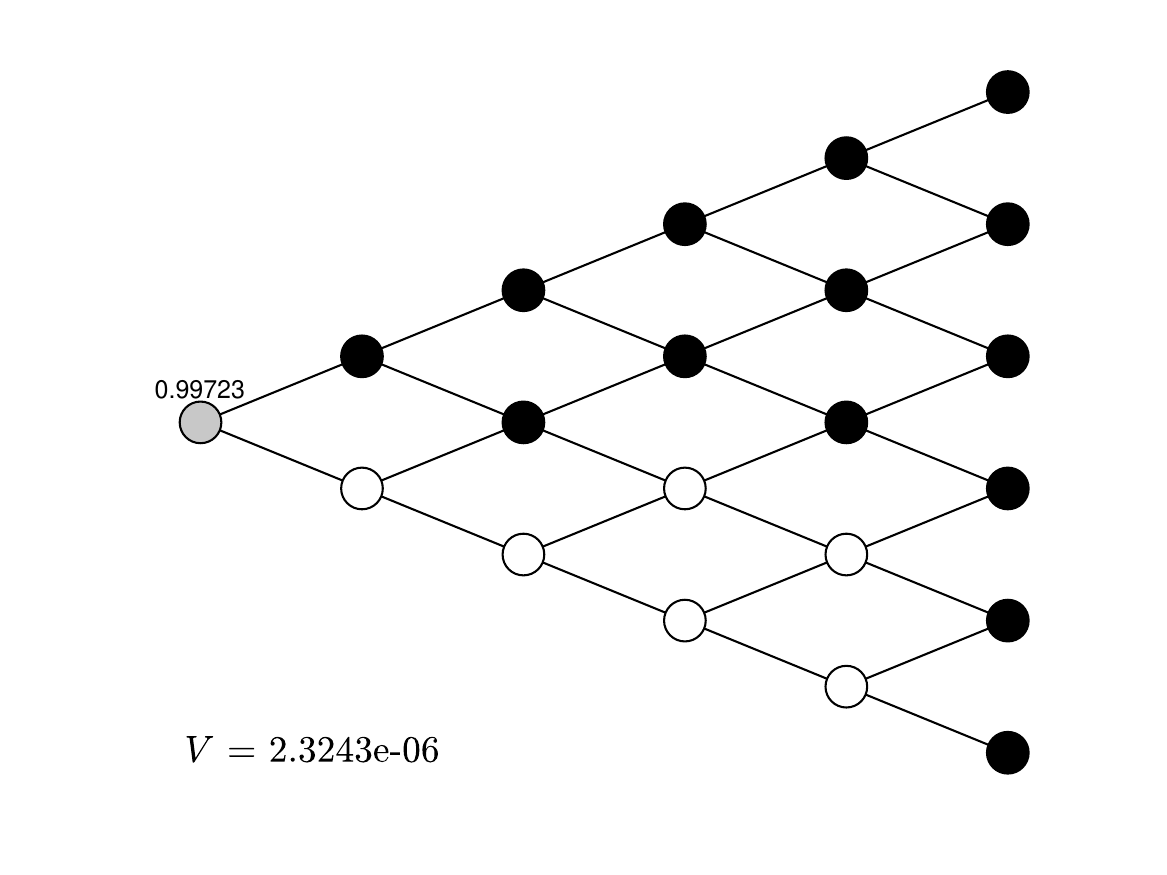}\\
  \end{minipage}
  \caption{Let $\alpha_\pm = 0.5$, $\delta_\pm = 0.9$, $\lambda = 1.5$. The naive strategy is turned into the sophisticated strategy after one round. The black nodes stand for stopping, the white nodes stand for continuing, and the grey nodes stand for randomization with the number above the node being the probability to stop.}\label{fig2}
\end{figure}

Next, let $\alpha_\pm = 0.5$, $\delta_\pm = 0.9$, $\lambda = 1.5$. The first graph of Figure \ref{fig2} shows the naive strategy under this group of parameter values, which is essentially stopping in gains and continuing in losses. The initial action is stopping at node $(0,0)$. After observing such actual behavior and taking the subsequent actions into consideration, the naive agent updates her strategy at each node, as shown in the second graph in Figure \ref{fig2}. The action at node $(0,0)$ is no longer stopping, but taking randomization with a large probability to stop. The naive strategy becomes exactly the same as the sophisticated one shown in the second graph after only one round of training. The corresponding objective value also becomes larger as the naive strategy approaches the sophisticated one.

Note that the degree of probability distortion is implied by the value of $\delta$. The smaller the $\delta$, the higher the degree of probability distortion. On the other hand, the time-inconsistency is due to the probability distortion. Then, the higher the degree of probability distortion, the more severe the level of time-inconsistency. It then takes more time to turn the naive strategies into the sophisticated ones when $\delta = 0.5$ than that when $\delta = 0.9$.

\subsection{Without randomization or arbitrary start}

\begin{figure}[!htbp]
  \centering
  \begin{minipage}[t]{0.327\textwidth}
  \centering
  \includegraphics[width=\textwidth]{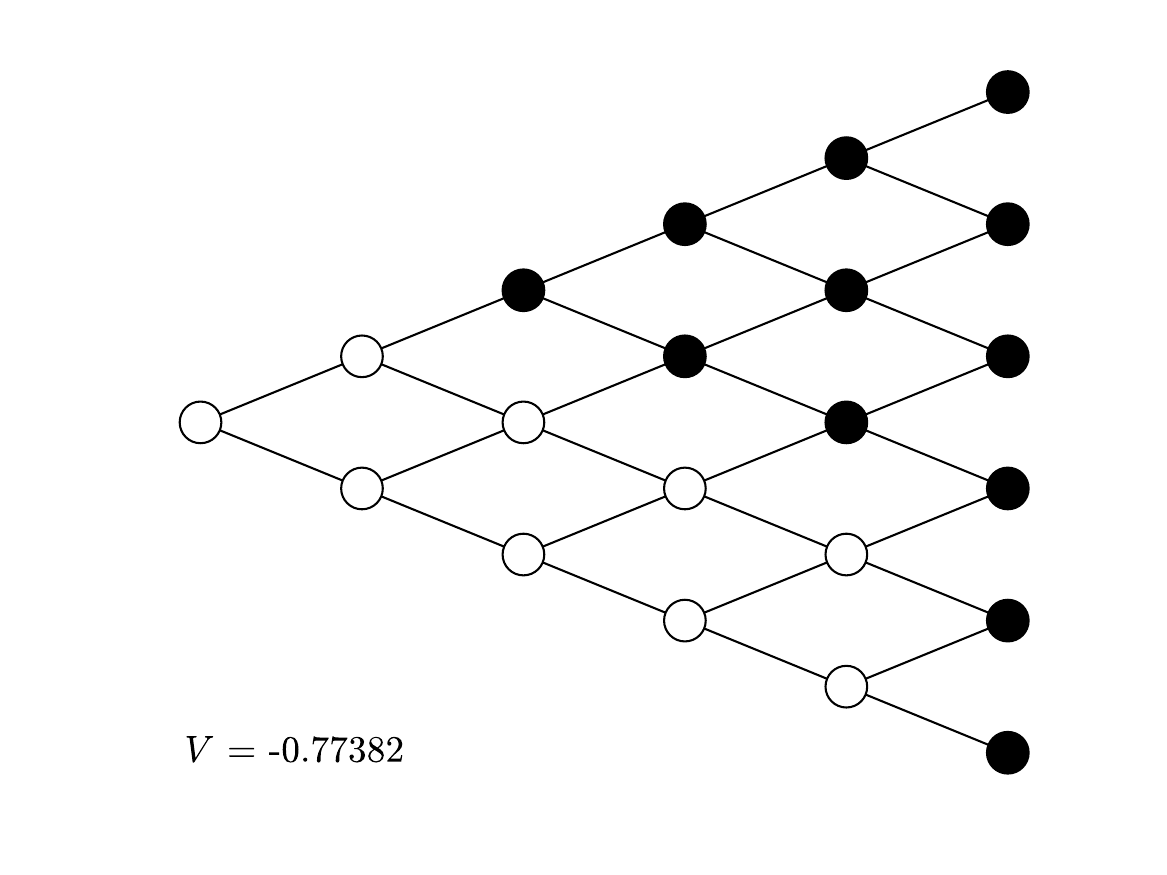}\\
  \end{minipage}
  \begin{minipage}[t]{0.327\textwidth}
  \centering
  \includegraphics[width=\textwidth]{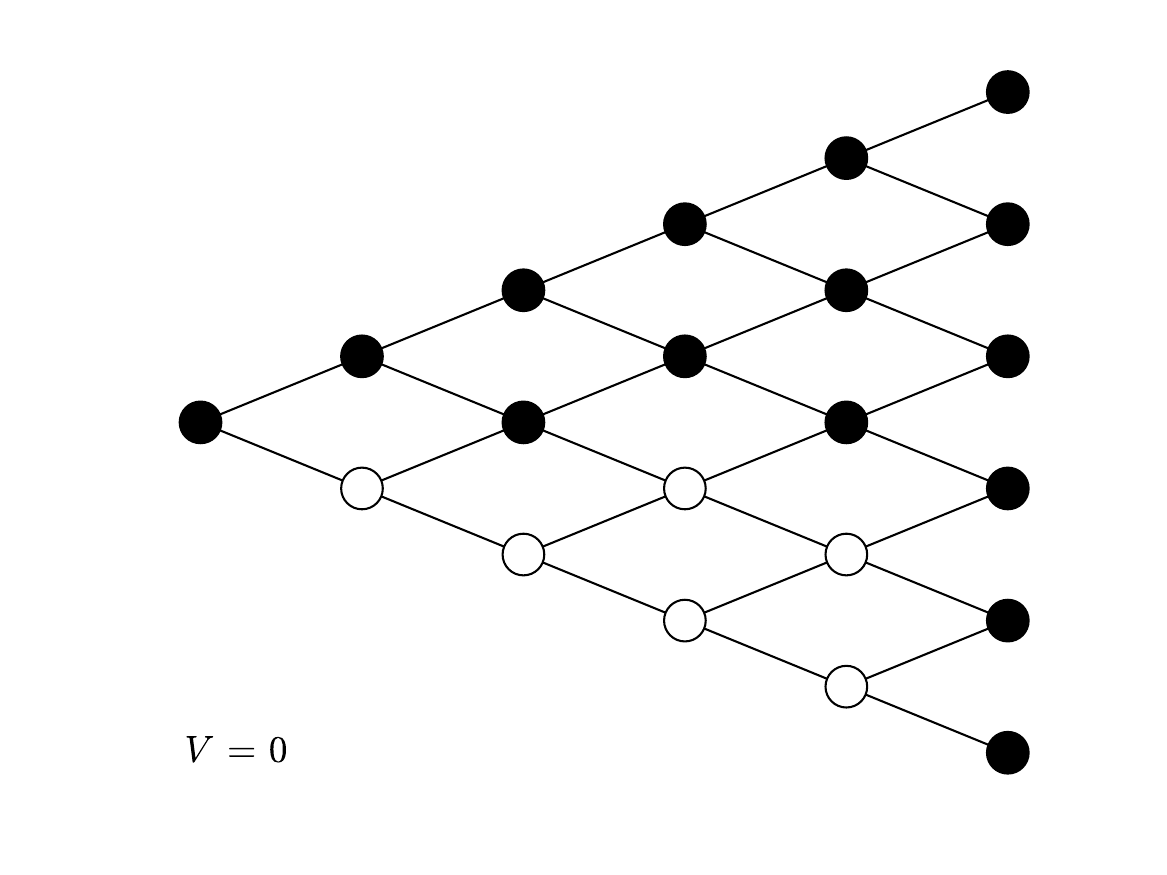}\\
  \end{minipage}
  \begin{minipage}[t]{0.327\textwidth}
  \centering
  \includegraphics[width=\textwidth]{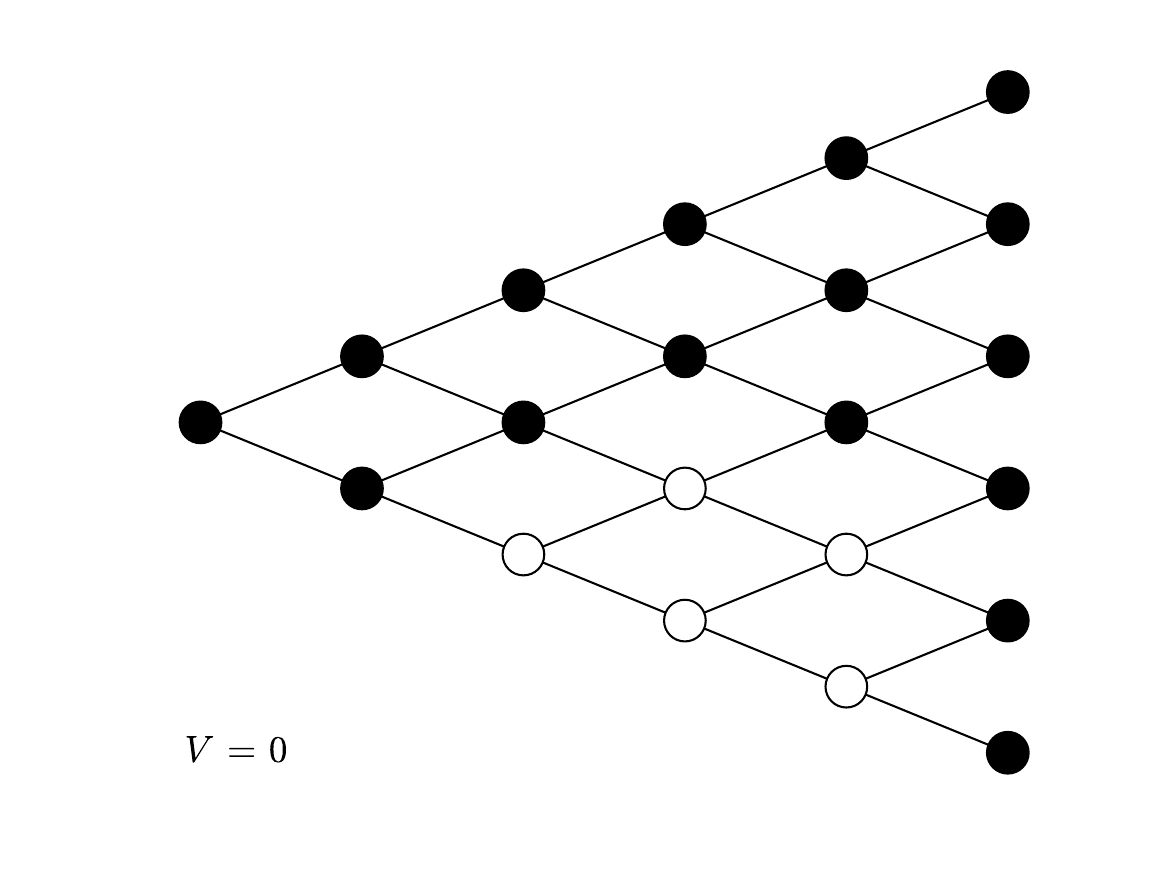}\\
  \end{minipage}
  \caption{Let $\alpha_\pm = 0.9$, $\delta_\pm = 0.5$, $\lambda = 1.5$. After two rounds the naive strategy without randomization is turned into the sophisticated strategy without randomization. The black nodes stand for stopping and the white nodes stand for continuing.}\label{fig1w}
\end{figure}

\begin{figure}[!htbp]
  \centering
  \begin{minipage}[t]{0.33\textwidth}
  \centering
  \includegraphics[width=\textwidth]{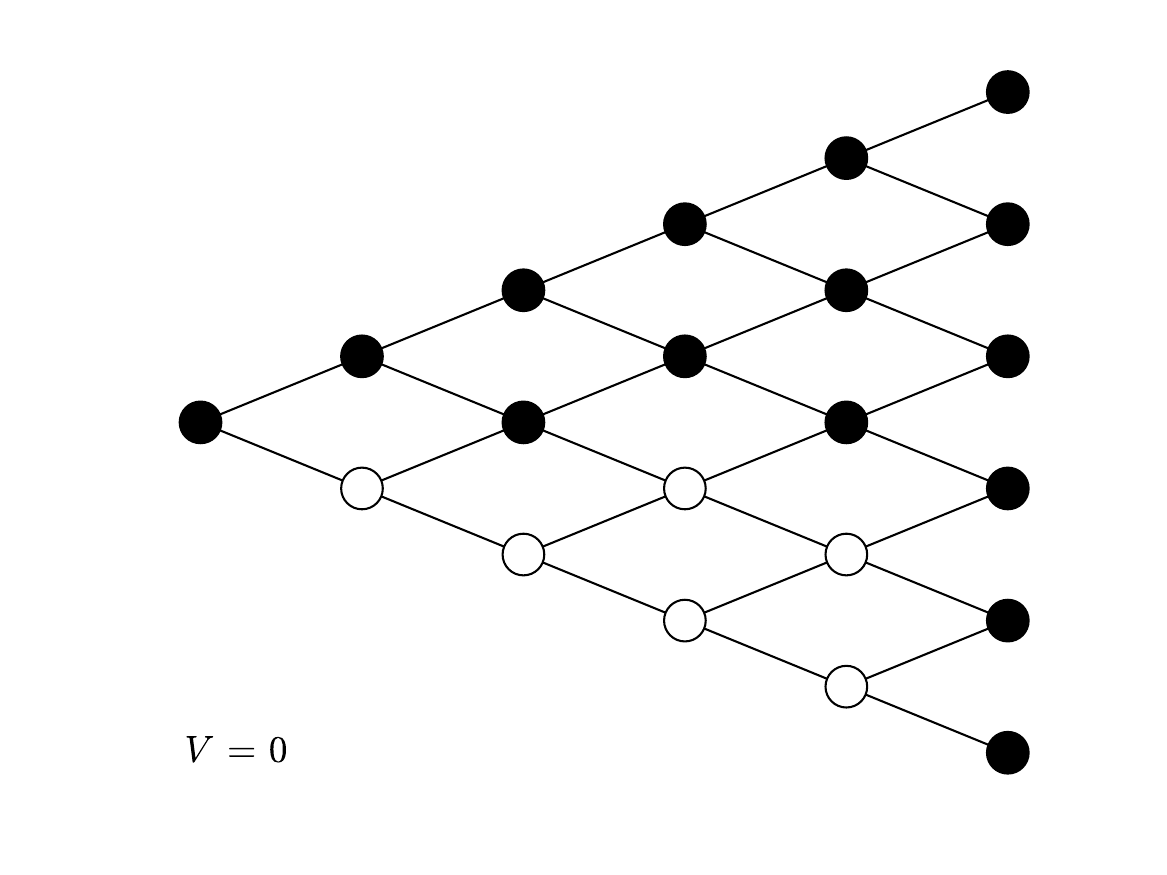}\\
  \end{minipage}
  \caption{Let $\alpha_\pm = 0.5$, $\delta_\pm = 0.9$, $\lambda = 1.5$. The naive strategy without randomization is the same as the sophisticated strategy without randomization. The black nodes stand for stopping and the white nodes stand for continuing.}\label{fig2w}
\end{figure}

We consider two extensions of the previous examples. First, we assume that the strategies are pure ones without randomization. This means that the agent can only choose a probability of 1 or 0 to be her action at each node. We find that the results are consistent with the previous ones with randomization. Figure \ref{fig1w} shows that for $\alpha_\pm = 0.9$, $\delta_\pm = 0.5$, $\lambda = 1.5$, after two rounds of training, the naive strategy is the same as the sophisticated one. Figure \ref{fig2w} shows that for $\alpha_\pm = 0.5$, $\delta_\pm = 0.9$, $\lambda = 1.5$, the naive strategy is exactly the same as the sophisticated one so no training is needed.

\begin{figure}[!htbp]
  \centering
  \begin{minipage}[t]{0.327\textwidth}
  \centering
  \includegraphics[width=\textwidth]{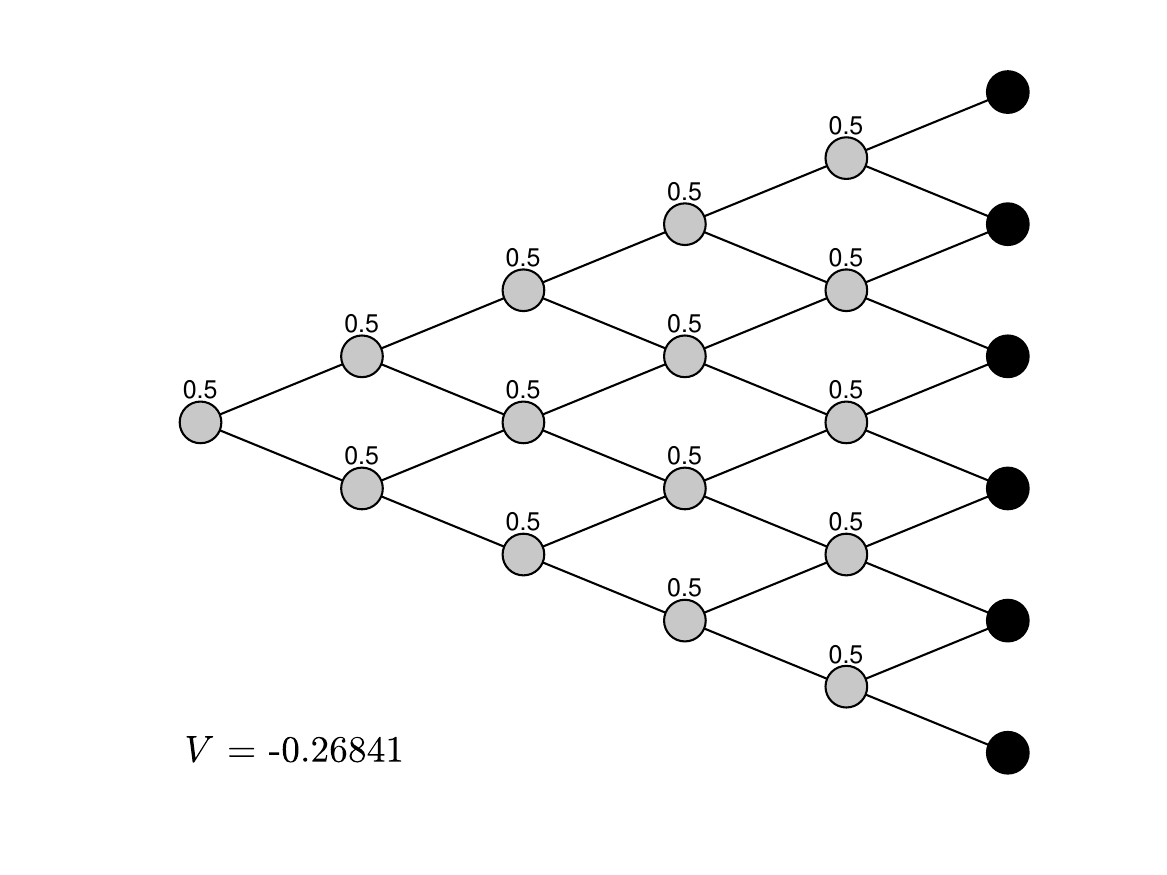}\\
  \end{minipage}
  \begin{minipage}[t]{0.327\textwidth}
  \centering
  \includegraphics[width=\textwidth]{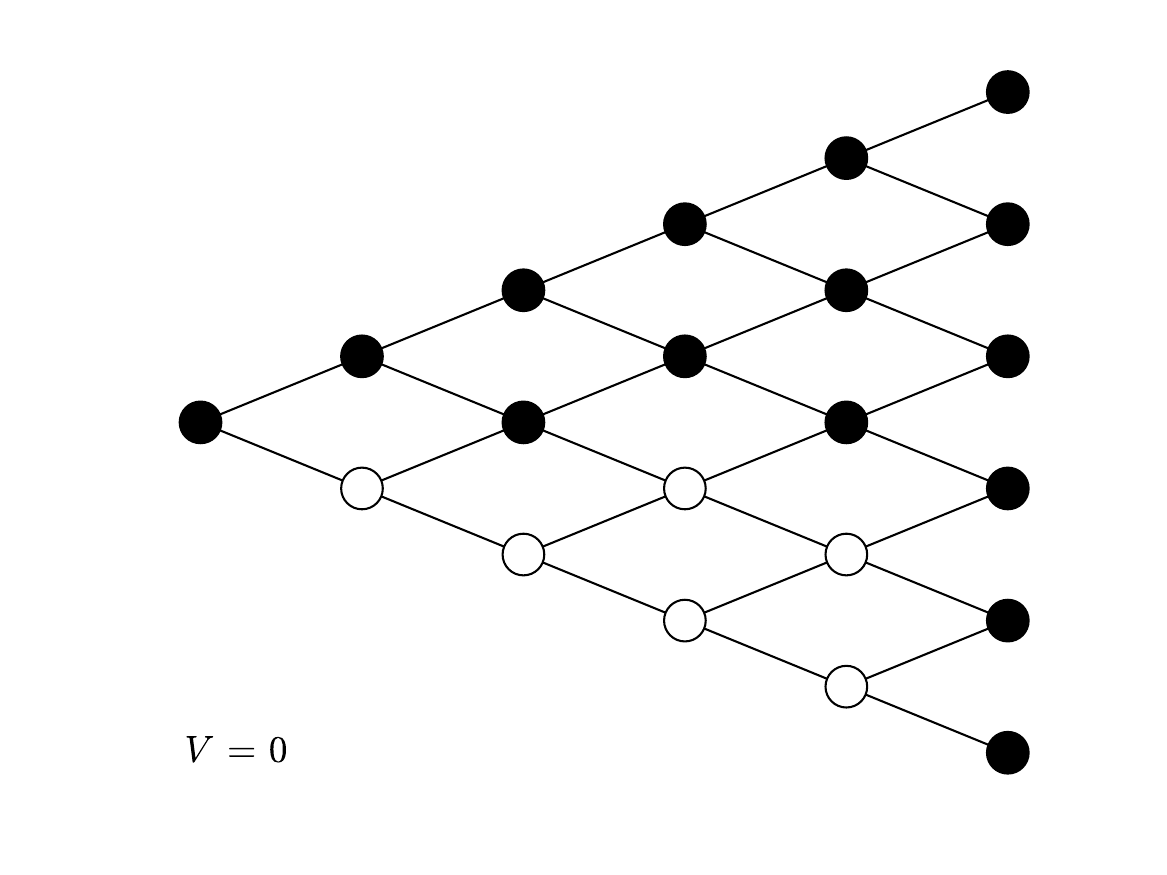}\\
  \end{minipage}
  \begin{minipage}[t]{0.327\textwidth}
  \centering
  \includegraphics[width=\textwidth]{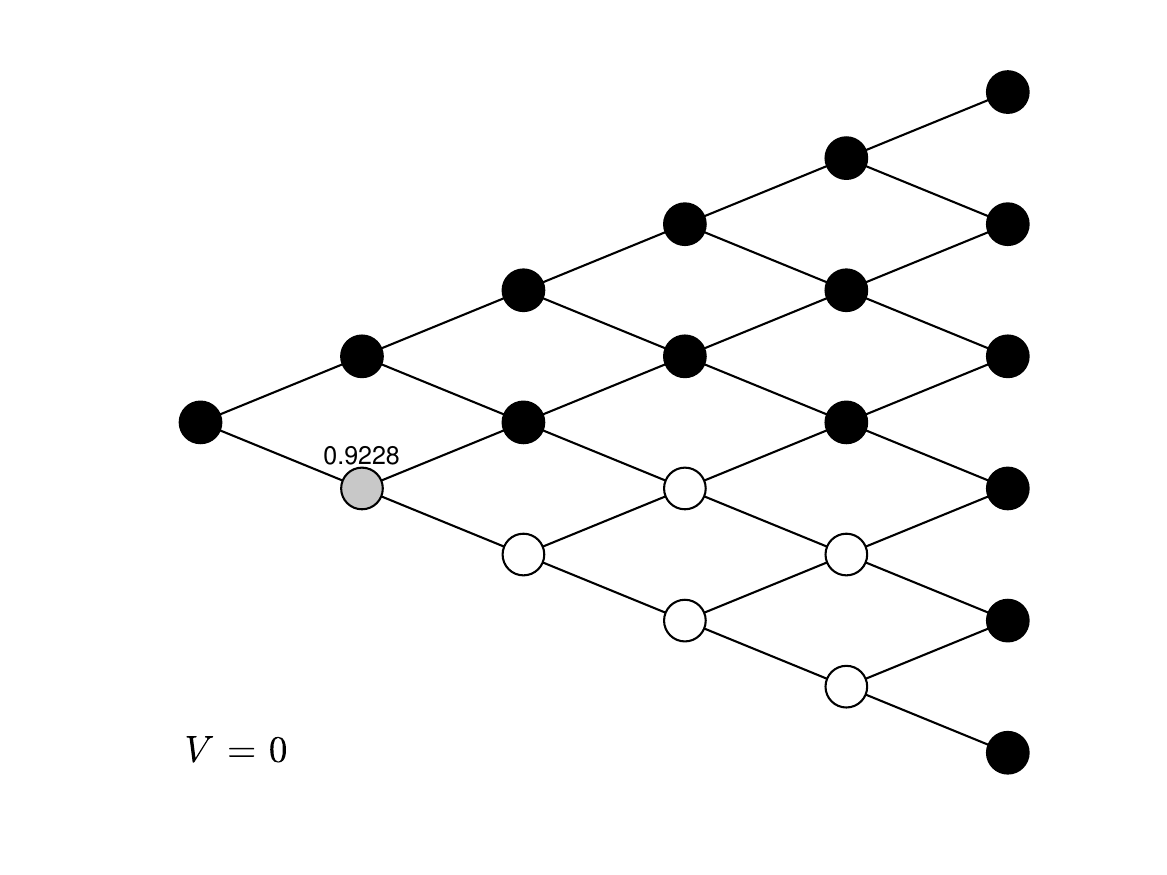}\\
  \end{minipage}
  \caption{Let $\alpha_\pm = 0.9$, $\delta_\pm = 0.5$, $\lambda = 1.5$. After two rounds the ``half-half'' strategy is turned into the sophisticated strategy. The black nodes stand for stopping, the white nodes stand for continuing, and the grey nodes stand for randomization with the number above being the probability to stop.}\label{fig1n}
\end{figure}

\begin{figure}[!htbp]
  \centering
  \begin{minipage}[t]{0.327\textwidth}
  \centering
  \includegraphics[width=\textwidth]{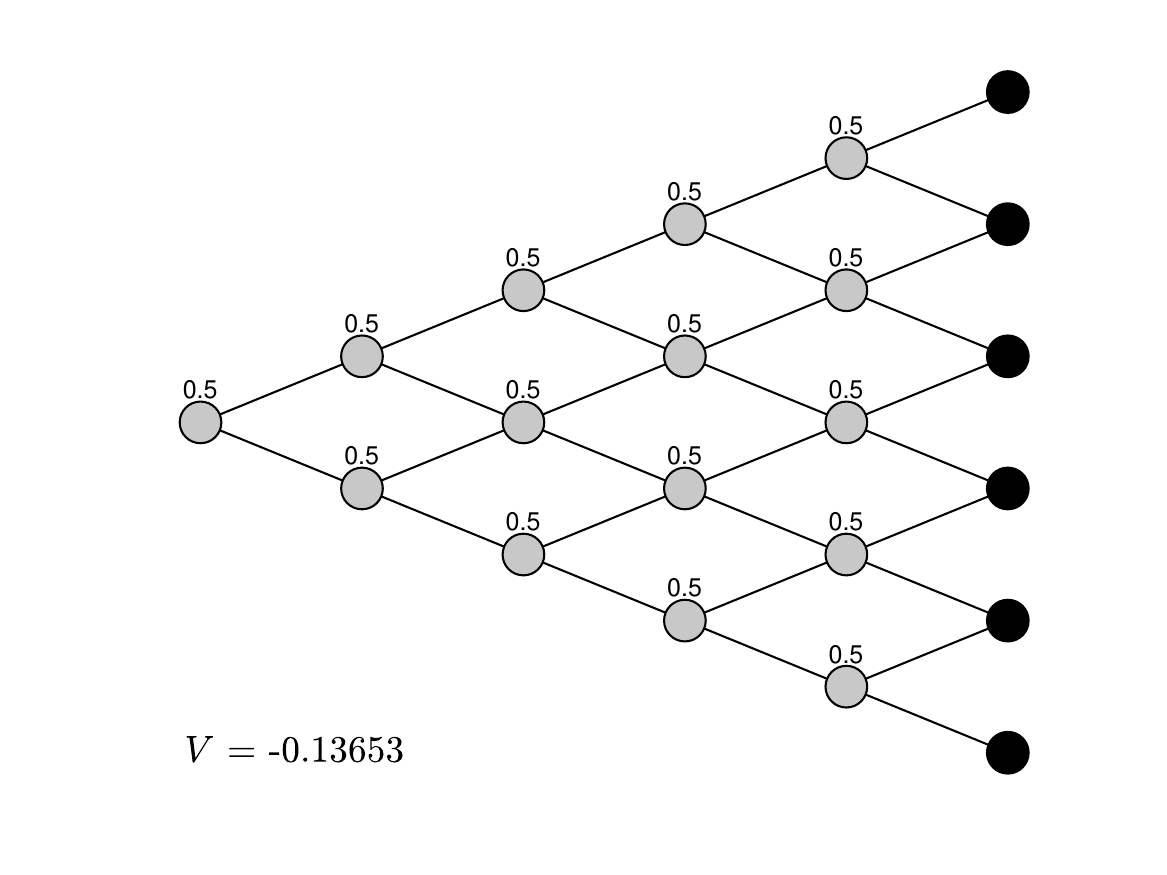}\\
  \end{minipage}
  \begin{minipage}[t]{0.327\textwidth}
  \centering
  \includegraphics[width=\textwidth]{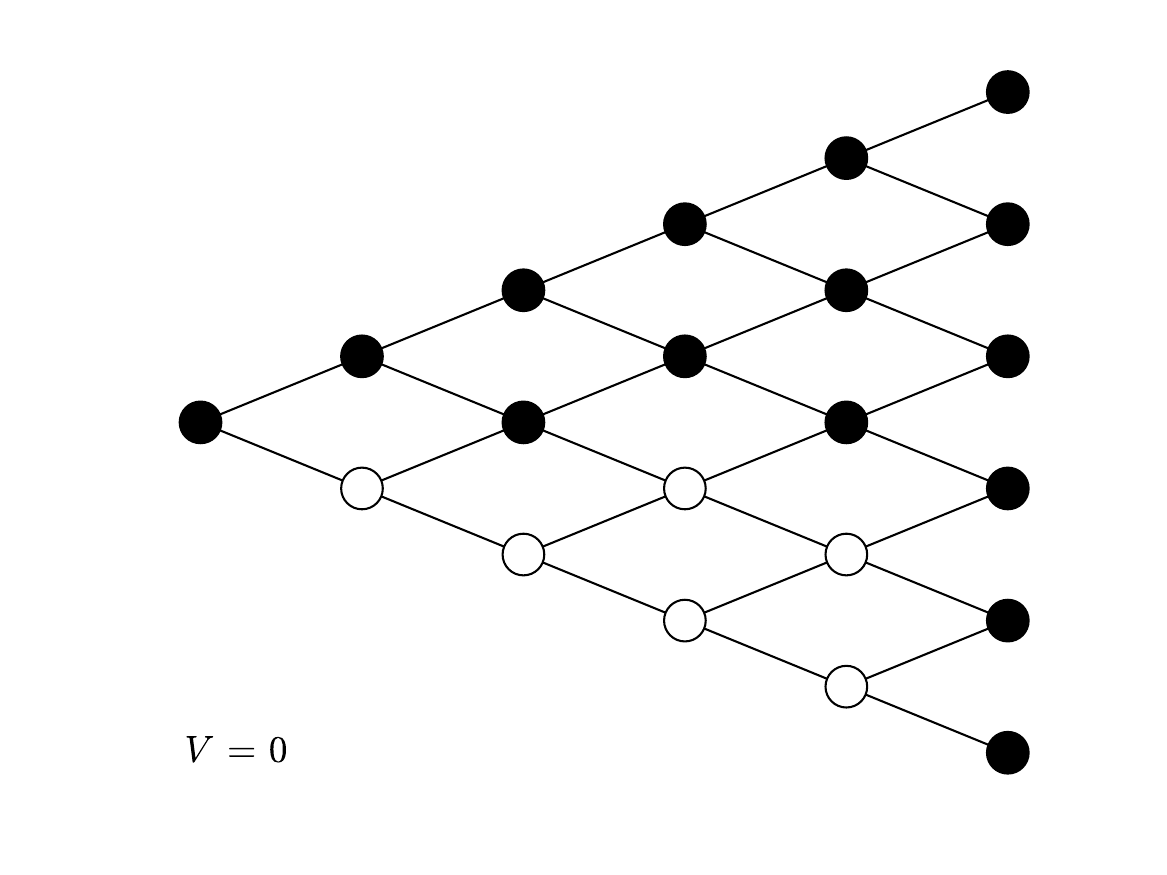}\\
  \end{minipage}
  \begin{minipage}[t]{0.327\textwidth}
  \centering
  \includegraphics[width=\textwidth]{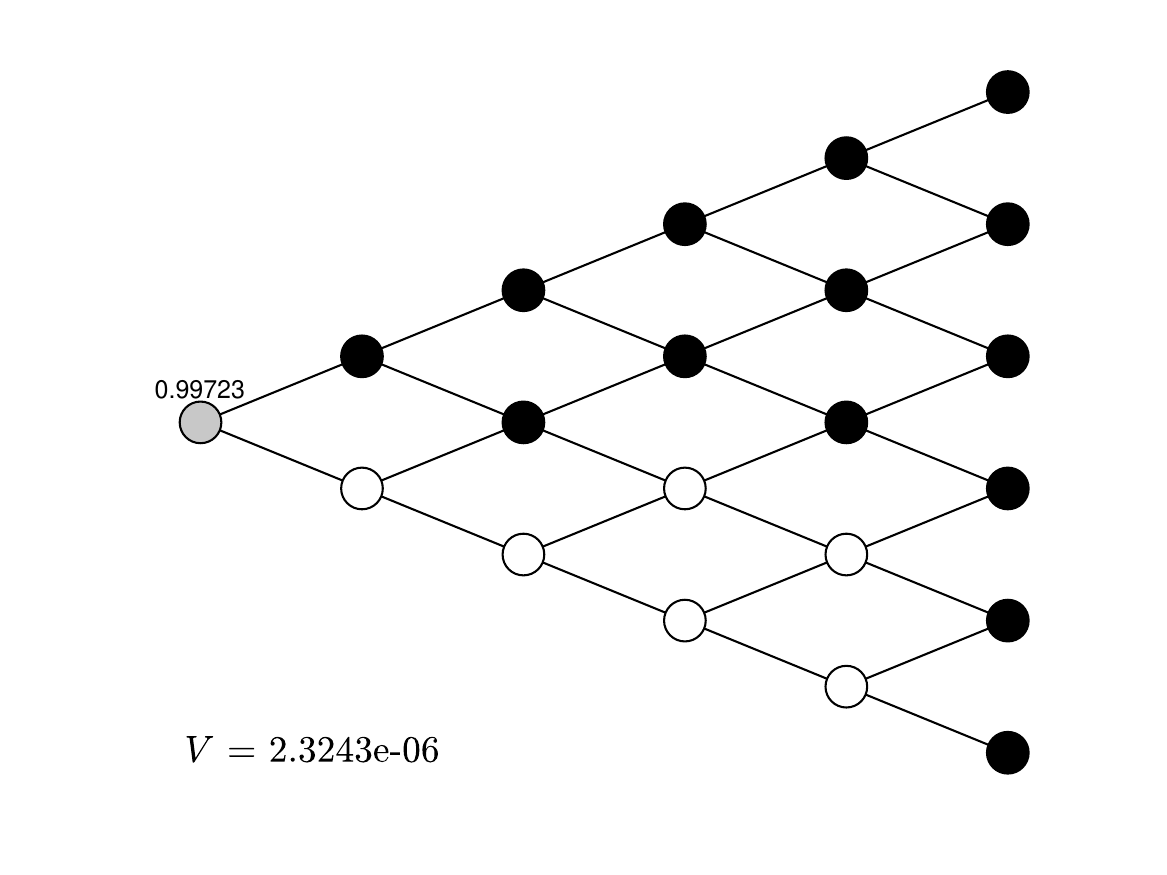}\\
  \end{minipage}
  \caption{Let $\alpha_\pm = 0.5$, $\delta_\pm = 0.9$, $\lambda = 1.5$. After two rounds the ``half-half'' strategy is turned into the sophisticated strategy. The black nodes stand for stopping, the white nodes stand for continuing, and the grey nodes stand for randomization with the number above being the probability to stop.}\label{fig2n}
\end{figure}

One can also start with an arbitrary strategy and then update it based on strategic reasoning, which eventually turns it into the sophisticated strategy. For $\alpha_\pm = 0.9$, $\delta_\pm = 0.5$, $\lambda = 1.5$, if one initially chooses the strategy as shown in the first graph of Figure \ref{fig1n} -- half the chance to continue and half the chance to stop -- after two rounds of training, this randomized strategy is turned into the sophisticated one, with an increasing CPT preference value. For $\alpha_\pm = 0.5$, $\delta_\pm = 0.9$, $\lambda = 1.5$, if one initially chooses the ``half-half'' strategy shown in the first graph of Figure \ref{fig2n}, then after two rounds of training this randomized strategy is also turned into the sophisticated one.

{\add

\section{Present-biased Preferences}\label{se:Presentbiased}

In this section, we consider the time-inconsistency problem due to present-biased preferences. Following \citet{ODonogheRabin:1999DoingItNowOrLater}, we consider two types of optimal stopping problem -- with immediate cost and with immediate reward -- and show that how naive strategies are turned into sophisticated strategies through a finite number of steps of reasoning.
 
\subsection{Immediate cost}

Suppose the optimal stopping problem with immediate cost is in $T$-period. If the agent stops immediately, the agent has an immediate cost $c$, while the reward $v$ is paid in the future and hence is discounted by a factor $\beta \in (0,1]$; if the agent stops later, both the cost and reward are discounted and the reward is reduced by $k$ amount proportionally. In other words, if the agent chooses to stop immediately, her preference value is $\beta v - c$. If the agent chooses to stop at time 1, her preference value is $\beta (v-k-c)$. If the agent chooses to stop at time 2, the preference value of such strategy perceived at time 0 is $\beta (v-2k-c)$, so on so forth. If the agent stops at terminal time $T$, her preference is $\beta (v-Tk-c)$. Suppose $k$ is small enough such that $k < v/T$ to make the reward always positive.

Note that the problem is state-independent, so the vertical nodes at the same time in the binomial tree can be collapsed into one. Consider time $t \in [0,T] \cap \mathbb{Z}$. Let $p_j$ be the probability of stopping at time $t+j$ from the perspective of time $t$, $j = 0,1,...T-t$. Then $p_0 + p_1 + ... + p_{T-t} = 1$. Let ${\bf a} = (p_0,p_1,...p_{T-t})$. Then the present-biased preference value of ${\bf a}$ at time $t$ is
\begin{align*}
V_{t}({\bf a}) = (\beta v - c) p_0 + \sum_{j = 1}^{T-t} \beta (v - j k - c)p_j - \beta tk.
\end{align*}

If $(1-\beta)c \leq \beta k$, it is always optimal for the naive agent to stop immediately at any time. Hence, 
\begin{align*}
{\bf a}^{N(0)} = \Big(\underbrace{1,... 1,...1}_{T+1}\Big),
\end{align*}
where 1 means that the agent stops immediately and 0 means that she continues for sure. It is straightforward to check that this naive strategy is equivalent to the sophisticated strategy, where the latter is derived through backward induction.

On the other hand, if $(1-\beta) c > \beta k$, it is never optimal for the naive agent to stop immediately due to the relatively large immediate cost $c$ until terminal time $T$. Hence, the naive strategy is to stop at the terminal time $T$:
\begin{align*}
{\bf a}^{N(0)} = \Big(\underbrace{0,... 0,...0}_{T},1\Big).
\end{align*}
Observing such stopping behavior, the naive agent would then update her strategy.
If further $(1-\beta) c \leq 2 \beta k$, then
\begin{align*}
{\bf a}^{N(1)} = \Big(\underbrace{1,... 1,...1}_{T-1},0,1\Big), \\
{\bf a}^{N(2)} = \Big(\underbrace{0,... 0,...0}_{T-2},1,0,1\Big),
\end{align*}
so on so forth, which eventually equal to the sophisticated strategy:
\begin{align*}
{\bf a}^{S} =
\begin{cases}
\Big(0,1,0,...1,0,1\Big) & \text{if $T$ is even} \\
\Big(1,0,1,... 1,0,1\Big) & \text{if $T$ is odd}.
\end{cases}
\end{align*}
In general, the following proposition shows how many steps are required to transform a purely naive strategy into a sophisticated one through strategic reasoning.
\begin{proposition}\label{prop:immediatecost}
Suppose $(1-\beta) c > \beta k$. Let $\varrho := \lceil (1-\beta)c / (\beta k) \rceil$. Then $\varrho \ge 2$.
The naive strategy is trained into the sophisticated one after $2(\lceil (T+1)/\varrho \rceil-1)$ rounds. 
\end{proposition}

\subsection{Immediate reward}
 
Suppose the optimal stopping problem with immediate reward is in $T$-horizon. If the agent stops immediately, the agent has an immediate reward $\theta^T v$, $\theta < 1$; if the agent stops one period later, the reward is increased by a degree of $\theta$. The discount factor $\beta \in (0,1]$. In other words, if she stops at time 0, the preference value is $\theta^T v$. If the agent stops at time 1, the preference value is $\beta \theta^{T-1} v $. If the agent chooses to stop at time 2, the preference value is $\beta \theta^{T-2} v$. If the agent chooses to stop at terminal time $T$, the preference value is $\beta v$.

Similar to the case of immediate cost, the stopping problem of immediate reward is also state-independent.
Consider time $t \in [0,T] \cap \mathbb{Z}$. Let $p_j$ be the probability of stopping at time $t+j$ from the perspective of time $t$, $j = 0,1,...T-t$. Then $p_0 + p_1 + ... + p_{T-t} = 1$. Let ${\bf a} = (p_0,p_1,...p_{T-t})$. Then the preference value of ${\bf a}$ at time $t$ is
\begin{align*}
V_{t}({\bf a}) = \left( \theta^T v p_0 + \sum_{j = 1}^{T-t} \beta \theta^{T-j} p_j \right) \theta^{-t}.
\end{align*}

If $\theta^T \ge \beta$, it is always optimal for the naive agent to stop immediately at any time. Hence,
\begin{align*}
{\bf a}^{N(0)} = \Big(\underbrace{1,... 1,...1}_{T+1}\Big),
\end{align*}
which is equivalent to the sophisticated strategy.
If $\theta < \beta$, it is optimal for the naive agent to stop as late as possible. Hence,
\begin{align*}
{\bf a}^{N(0)} = \Big(\underbrace{0,... 0,...0}_{T},1\Big),
\end{align*}
which is also equivalent to the sophisticated strategy. The following proposition shows the remaining cases of transformation from naive strategy to sophisticated one.
\begin{proposition}\label{prop:immediatereward}
Suppose $\theta \ge \beta > \theta^T$. Let $\nu := \lfloor \log \beta / \log \theta \rfloor $. Then $\nu \in [1,T-1] \cap \mathbb{Z}$. For $\nu \ge 1$, the naive strategy is turned into the sophisticated one after $\lceil T/\nu \rceil$ rounds.
\end{proposition}

}

\section{Conclusion}\label{se:Conclusion}

We consider the optimal stopping problem with time-inconsistency in a discrete-time setting with randomization allowed. Because of time-inconsistency, the naive agent deviates from any of her predetermined plans through optimizing her action at each time-state node, whereas the sophisticated agent plans a consistent strategy by taking her future selves actions into consideration. When the naive agent can observe her actual behavior and take her subsequent real actions into consideration, her strategies eventually match with sophisticated strategies after several rounds of training. Under the cumulative prospect theory preferences where the time-inconsistency is due to the probability distortion, the higher the degree of probability distortion, the more time required to turn the naive strategies into the sophisticated ones and hence the more severe the level of time-inconsistency. For strategies without randomization or arbitrary initial strategies, the same algorithm can be applied to turn them into the sophisticated strategies. {\add For time-inconsistency due to the present-biased preferences, we provide analytical results on transforming the time-inconsistent strategies into time-consistent ones in optimal stopping problem with immediate cost and with immediate reward.} The analysis shows that strategic reasoning is powerful in achieving time-consistent plans in time-inconsistent problems.

\vspace{20ex}

\appendix
\begin{appendices}

\section{Appendix}
\pfof{Proposition \ref{prop:Tround}}
For a $T$-period binomial tree, the actions taken at the terminal time $T$ should be 1 because the agent should stop at time $T$ if she has not stopped yet. Therefore, for both the naive and the sophisticated agents, $a^{N(0)}(T,x) = a^{S}(T,x) = 1$, $x \in \{-T,-(T-2),...,T-2,T\}$. Meanwhile, the naive agents' actions at time $T-1$ is $a^{N(0)}(T-1,x) = p^*$, which is obtained by optimizing $V_{T-1,x}({\bf a})$, where ${\bf a} = (p,a(T,x+1),a(T,x-1))$, subject to $p \in [0,1]$ and $a(T,x+1) = a(T,x-1) = 1$.
This is exactly same to the sophisticated agent's action planned at time $T-1$, that is, $a^{N(0)}(T-1,x) = a^{S}(T-1,x)$, for $x \in \{-(T-1),-(T-3),...,T-3,T-1\}$.
After the first round of training, the naive agent's actions taken at time $T-2$ become $a^{N(1)}(T-2,X_{T-2}) = p^*$, which is obtained by optimizing $V_{T-2,X_{T-2}}({\bf a})$, where ${\bf a} = (p,a(T,x+1),a(T,x-1))$, subject to $p \in [0,1]$, $a(T-1,X_{T-1}) = a^{N(0)}(T-1,X_{T-1})$, and $a(T,X_T) = a^{N(0)}(T,X_T)$.
This is exactly same to the sophisticated agent's action planned at time $T-2$, that is, $a^{N(1)}(T-2,X_{T-2}) = a^{S}(T-2,X_{T-2})$.
With the same logic, one obtains that $a^{N(k-1)}(T-k,X_{T-k}) = a^{S}(T-k,X_{T-k})$, where $k = 1,2,...T$. Then, after $T-1$ rounds, $a^{N(T-1)}(t,x)=a^{S}(t,x)$ for any node $(t,x)$.
\qed

{\add

\pfof{Proposition \ref{prop:immediatecost}}
If $\varrho > T$, then $(1-\beta) c > T \beta k$. In this case, realizing that herself is going to stop at the terminal time $T$ according to ${\bf a}^{N(0)}$, the naive agent finds that doing so is indeed optimal. Then ${\bf a}^{N(1)}$ is exactly same as ${\bf a}^{N(0)}$, which means that the naive strategy is same as the sophisticated one.

If $\varrho \in [2,T] \cap \mathbb{Z}$, then $(\varrho-1) \beta k < (1-\beta) c \le \varrho \beta k$. Then comparing the preference value of stopping at the terminal time $T$ according to ${\bf a}^{N(0)}$ with the value of stopping immediately at time $t < T$, the naive agent updates her strategy to be
\begin{align*}
{\bf a}^{N(1)} = \Big(\underbrace{1,...1}_{T-\varrho+1},\underbrace{0,...0}_{\varrho-1},1\Big).
\end{align*}
Similarly, her strategy through another round of strategic reasoning becomes
\begin{align*}
{\bf a}^{N(2)} = \Big(\underbrace{0,...0}_{T-\varrho},1,\underbrace{0,...0}_{\varrho-1},1\Big).
\end{align*}
If $T-\varrho + 1 \le \varrho$, then no further change is made and ${\bf a}^{N(3)}$ is same as ${\bf a}^{N(2)}$, which means that it has become the sophisticated strategy.
If $T-\varrho + 1 > \varrho$, then the third-round of strategic reasoning leads to 
\begin{align*}
{\bf a}^{N(3)} = \Big(\underbrace{1,...1}_{T-2\varrho},1,\underbrace{0,...0}_{\varrho-1},1,\underbrace{0,...0}_{\varrho-1},1\Big),
\end{align*}
so on so forth, until it is turned into the sophisticated strategy:
\begin{align*}
{\bf a}^{S} = \Big(\underbrace{0,...0}_{T-(\omega-1) \varrho},1,\underbrace{\underbrace{0,...0}_{\varrho-1},1,...\underbrace{0,...0}_{\varrho-1},1}_{(\omega-1)\varrho} \Big),
\end{align*}
where $\omega := \lceil (T+1)/\varrho \rceil$. In total, it takes $2(\omega-1)$ rounds to achieve ${\bf a}^{S}$.
\qed

\pfof{Proposition \ref{prop:immediatereward}}
Note that according to the definition of $\nu$, we have $\theta^{\nu} \ge \beta > \theta^{\nu+1}$.
Then it is not optimal for the naive agent to stop immediately until the time horizon is no longer than $\nu$, that is,
\begin{align*}
{\bf a}^{N(0)} = \Big(\underbrace{0,... 0}_{T-\nu},\underbrace{1,... 1}_{\nu},1\Big).
\end{align*}
If $T-\nu \le \nu$,
\begin{align*}
{\bf a}^{N(1)} = \Big(\underbrace{1,...1,...1}_{T+1}\Big),
\end{align*}
which is equivalent to the sophisticate strategy.
If $T-\nu > \nu$,
\begin{align*}
{\bf a}^{N(1)} = \Big(\underbrace{0,... 0}_{T-2\nu},\underbrace{1,... 1}_{2\nu},1\Big),
\end{align*}
so on so forth, until it is equivalent to the sophisticate strategy:
\begin{align*}
{\bf a}^{S} = \Big(\underbrace{1,...1,...1}_{T+1}\Big).
\end{align*}
In total, it takes $\lceil T/\nu \rceil$ rounds from the naive strategy to the sophisticated strategy.
\qed
}

\end{appendices}

\bibliography{LongTitles,BibFile}

\end{document}